\documentclass[amsmath,amssymb,reprint,
 aps]{revtex4-1}
\usepackage{xcolor,comment}
\usepackage{graphicx}% Include figure files
\usepackage{dcolumn}% Align table columns on decimal point
\usepackage{bm}% bold math
\usepackage{xcolor}
\usepackage{hyperref}% add hypertext capabilities
\newcommand{\be}{\begin{equation}}
\newcommand{\ee}{\end{equation}}
\newcommand{\bea}{\setlength\arraycolsep{2pt} \begin{eqnarray}}
\newcommand{\eea}{\end{eqnarray}}
\newcommand{\nn}{\nonumber}

\def\ft#1#2{{\textstyle{\frac{\scriptstyle #1}{\scriptstyle #2} } }}
\def\fft#1#2{{\frac{#1}{#2}}}
\begin{document}

\title{Higher-order correction to weak-field lensing of an Ellis-Bronnikov wormhole}% Force line breaks with \\

\author{Tingqi Cai}
\email{tcaiac@connect.ust.hk}
\affiliation{Department of Physics, The Hong Kong University of Science and Technology, Clear Water Bay, Kowloon, Hong Kong, People’s Republic of
China}
\affiliation{Jockey Club Institute for Advanced Study, The Hong Kong University of Science and Technology, Clear Water Bay, Kowloon, Hong Kong,
People’s Republic of China}
\author{Hyat Huang}
\email{hyat@mail.bnu.edu.cn}
\affiliation{College of Physics and Communication Electronics, Jiangxi Normal University, \\
Nanchang 330022, China and\\
Institute of Physics, University of Oldenburg, 
\\Postfach 2503, D-26111 Oldenburg,
Germany
}%
\author{Zun Wang}
\email{zwangdq@connect.ust.hk}
\affiliation{Department of Physics, The Hong Kong University of Science and Technology, Clear Water Bay, Kowloon, Hong Kong, People’s Republic of
China}
\affiliation{Jockey Club Institute for Advanced Study, The Hong Kong University of Science and Technology, Clear Water Bay, Kowloon, Hong Kong,
People’s Republic of China}

\author{Mian Zhu}
\email{mzhuan@connect.ust.hk}
\affiliation{Faculty of Physics, Astronomy and Applied Computer Science, Jagiellonian University, 30-348 Krakow, Poland}

\date{\today}% It is always \today, today,

\begin{abstract}
The gravitational lensing effect at higher order under weak-field approximation is believed to be important to distinguish black holes and other compact objects such as wormholes. The deflection angle of a generic wormhole is difficult to solve analytically; thus approximation methods are implemented. In this paper, we investigate the weak-field deflection angle of a specific wormhole, the Ellis-Bronnikov wormhole, up to the $1/b^4$ order. We use different approximation formalisms, study their precision at $1/b^4$ order by a comparison to a purely numerical result, and finally rank these formalisms by their accuracy. Moreover, we find that certain formalisms are sensitive to the choice of coordinate system; thus it is important to choose the coordinate system appropriately for the evaluating of lensing physics.

\end{abstract}

\maketitle

\section{Introduction}
Gravitational lensing is an important probe to study the physics of compact objects \cite{refsdal1994gravitational,Wambsganss:1998gg}. The recent astrophysical discovery, such as the observation of gravitational waves from compact objects \cite{LIGOScientific:2016aoc}, made it possible to detect compact objects through lensing physics in the near future. Thus, there is a growing interest in the study of gravitational lensing, and some up-to-date work can be found in \cite{Tsukamoto:2021lpm,Ren:2021uqb,Qiao:2021trw,Javed:2022psa,Liu:2022lfb,Atamurotov:2022knb,Gao:2022cds,Luo:2022uij,Ghosh:2022mka,Sengo:2022jif,Guo:2022muy,Qiao:2022nic,Huang:2023iog,Shan:2023ngi,Suyu:2023jue,AbhishekChowdhuri:2023ekr}. 

The lensing effect is extensively studied for different lenses \cite{Dabrowski:1998ac,Safonova:2001vz,Eiroa:2002mk,Virbhadra:2002ju,TejeiroS:2005ltc,Nandi:2006ds}. Unfortunately, it turns out to be difficult to distinguish black holes (BHs) and other compact objects such as wormholes (WHs)\cite{Morris:1988cz,Morris:1988tu} and boson stars \cite{Schunck:2003kk}. Those compact objects can mimic the behavior of a black hole \cite{Abramowicz:2002vt,Damour:2007ap,Guzman:2009zz,Tsukamoto:2012xs, Abdikamalov:2019ztb}, and we are unable to observationally exclude their existence at the present stage \cite{Berti:2015itd,Barack:2018yly, Cardoso:2019rvt,LISA:2022kgy}. In view of the theoretical importance of black holes as well as other compact objects, it is essential to search for distinctive features of these compact objects \cite{Harko:2009gc,Kovacs:2010xm,Sahu:2012er,Konoplya:2016hmd,Cunha:2017wao,Shaikh:2019hbm,Karimov:2020fuj,Bronnikov:2021liv,Vagnozzi:2022moj}.

There are two motivations for us to consider the lensing effect from a wormhole. On the one hand, recent developments on the light ring reveal unavoidable instabilities for a large variety of horizonless compact objects (and thus exclude the possibility for them to be astrophysically observed), while wormholes might be free from these instabilities \cite{Cardoso:2014sna,Cunha:2017qtt,Cunha:2022gde}. On the other hand, lensing effect from higher-order contributions might be eligible to distinguish black holes and their correspondent wormholes in the weak-field limit \cite{Bronnikov:2018nub,Izmailov:2020ypf}. However, it is generically difficult to analytically resolve the deflection angle for lenses with a complicated metric. Approximation methods are then developed to simplify the evaluation procedure, for example, the Amore-Diaz (AD) formalism \cite{Amore:2006pi}, the formalism through Gauss-Bonnet theorem (GBT) \cite{Gibbons:2008rj}, and the post-post-Newtonian (PPN) methods developed by Keeton and Petters \cite{Keeton:2005jd}. While these methods yield rather accurate results at leading order, their predictions on higher-order contributions are different from each other. Unfortunately, recent research pointed out that contributions from higher-order terms are important to distinguish different lens objects, for example, the black holes and their mimicker (see, e.g., \cite{Izmailov:2020ypf}). Thus, it is important to clarify the precision of the gravitational lensing effect with different approximation methods.

In view of the above argument, we study the higher-order gravitational lensing effect with different approximation methods, in an exemplified wormhole. One of the most well-known wormhole solutions is the Ellis-Bronnikov wormhole (EBWH), discovered in 1973 \cite{Ellis:1973yv,Ellis:1979bh,Bronnikov:1973fh}. It is based on the theory of Einstein gravity coupled to a free scalar field, which is referred to as a “phantom field." This field is defined by flapping the sign of its kinetic term in the Lagrangian. It was originally believed that this phantom field causes instability in the EBWH, but a recent study has argued that the instability can be cured \cite{Piazza:2004df}. Additionally, another study suggests that slow rotation can stabilize the EBWH \cite{Azad:2023iju}. As a simple yet significant model of a traversable wormhole, the gravitating lensing effect of the EBWH has garnered widespread attention in the scientific community. Most studies on the lensing effect have focused on a special case of the EBWH, where the wormhole mass is zero \cite{Abe:2010ap,Toki:2011zu,Nakajima:2012pu,Yoo:2013cia,Bozza:2015haa,Lukmanova:2016czn,Tsukamoto:2016qro,Jusufi:2017gyu,Asada:2017vxl,Tsukamoto:2017hva,Bhattacharya:2019kkb}. 

In this work, we extend the study to the full EBWH metric and evaluate the deflection angle up to the $1/b^4_\pm$ order in the weak-field limit ($b_\pm$ is the rescaled impact parameter). We find that the PPN result is the most accurate one. When the weak-field condition is better satisfied, the GBT formalism is more precise than the AD formalism. When the weak-field limit is less satisfied, the AD formalism might yied better accuracy; however, in this case the formalism based on weak field approximation might not be applicable. Moreover, we find that for AD and GBT formalisms, the results are sensitive to the choice of coordinate system when the corresponding deflection angle approaches $0$ in the negative-mass branch of universe.

The paper is organized as follows. We discuss the Ellis-Bronnikov wormhole in Sec. \ref{sec:EBWH}, then briefly introduce the lensing physics in Sec. \ref{sec:lensing}. In Sec. \ref{sec:massless}, we make a pedagogical introduction to the approximation formalism, using the massless EBWH as an example. We present our main result, the deflection angle for general EBWH up to second order in different formalisms, in Sec. \ref{sec:massive}, and discuss the precision of different methods. We conclude in Sec. \ref{sec:conclusion}.

Throughout this paper, we will adopt the Planck units so that $8\pi G=c=1$. We use the $(-,+,+,+)$ convention. 

\section{Brief introduction to Ellis-Bronnikov wormhole}
\label{sec:EBWH}

\subsection{Theory and metric}

One of the simplest ways to construct a wormhole solution is by introducing a free phantom scalar field as exotic matter in general relativity, namely
\be
{\cal L}= \sqrt{-g}(R+\ft{1}{2}(\partial \phi)^2),
\ee
where $R$ is the scalar curvature and $\phi$ is a phantom scalar field. The EBWH was discovered independently by Ellis and Bronnikov in 1973 and is a spherically symmetric solution. The line element of EBWH can be written as 
\bea\label{bronnikov}
&&ds^2=-h(r) dt^2+ h(r)^{-1}dr^2+R^2(r)d\Omega_{2}^2\,,\nn\\
&&h=e^{-\fft{m}{q}\phi},\qquad R^2=\fft{r^2+q^2-m^2}{h}\,,\nn\\
&&\phi=\fft{2q}{\sqrt{q^2-m^2}}\arctan(\fft{r}{\sqrt{q^2-m^2}})\,,\label{eq:ebsol}
\eea
where $(m,q)$ are two integration constants. Here $d\Omega_2^2$ represents the unit solid angle for two dimensions. The wormhole throat, which connects two asymptotic flat spacetime regions, is located at the minimum of $R(r)$ and is at $r=-m$. These regions can be referred to as Universe I where $r\in(-\infty, -m)$ and Universe II where $r\in(-m, +\infty)$. When $m\neq 0$, the metric denotes an asymmetric wormhole because Universe I is not a copy of Universe II. 

Specifically, the two universe would observe the wormhole with opposite signs and different absolute values:
\begin{equation}
\label{eq:Mpm}
    M_{\pm} = \pm m e^{ \pm \pi m/(2\sqrt{q^2-m^2})} ~.
\end{equation}
If $m\geq 0$ and $q\geq 0$, Universe I has a positive wormhole mass while Universe II has a negative one. For more details on the global structure of EBWH, refer to Ref.\cite{Huang:2020qmn}.

Later on, we may simply use the $\pm$ sign to refer to the universe with an observed positive/negative mass, if convenient. Moreover, we see the gravitational property of the wormhole seen by Universe II is equivalent to that by Universe I, as long as we make a transformation $m \to -m$. Thus, we may set $m \geq 0$ without loss of generality. Although the two universes are asymmetric, in our convention \eqref{eq:Mpm} they are written as symmetric as possible, differing only by a transform $m \to -m$.

We mention that, the horizonless condition for EBWH, i.e. the metric component $g_{00}$ is everywhere negative, is imposed by $q^2 > m^2$. Since we have set $m \geq 0$, the condition can be simply written as $q > m$. 

\subsection{Special cases}
\label{sec:cases}
When $m=0$, the metric \eqref{bronnikov} reduces to the well-known symmetric EB wormhole:
\be
ds^2=-dt^2+dr^2+(r^2+q^2)d\Omega^2_2,
\ee
where the light ring locates at $r=0$ and without ISCO in the whole spacetime.

In principle, the horizonless condition $q^2 > m^2$ forbids the possible $q \to 0$ limit. However, if we naively set $q=0$ in the metric \eqref{eq:ebsol}, we may recover the Schwarzschild black hole with mass $m$ (see, e.g., \cite{Huang:2020qmn} for more details). This is in agreement with our intuition: the condition $q^2 > m^2$ forbids the existence of a horizon, so when we take $q=0$ and $m \neq 0$, we expect a black hole to come to exist. 

\section{Basics on gravitational lensing physics}
\label{sec:lensing}
\subsection{Lensing geometry}
For simplicity, we start with a static spherically symmetric metric
\begin{equation}
\label{eq:SSSmetric}
    ds^2 = -P(r) dt^2 + Q(r) dr^2 + R^2(r) d\Omega_2^2 ~.
\end{equation}

\begin{figure}[ht]
    \centering
    \includegraphics[width=0.96\linewidth]{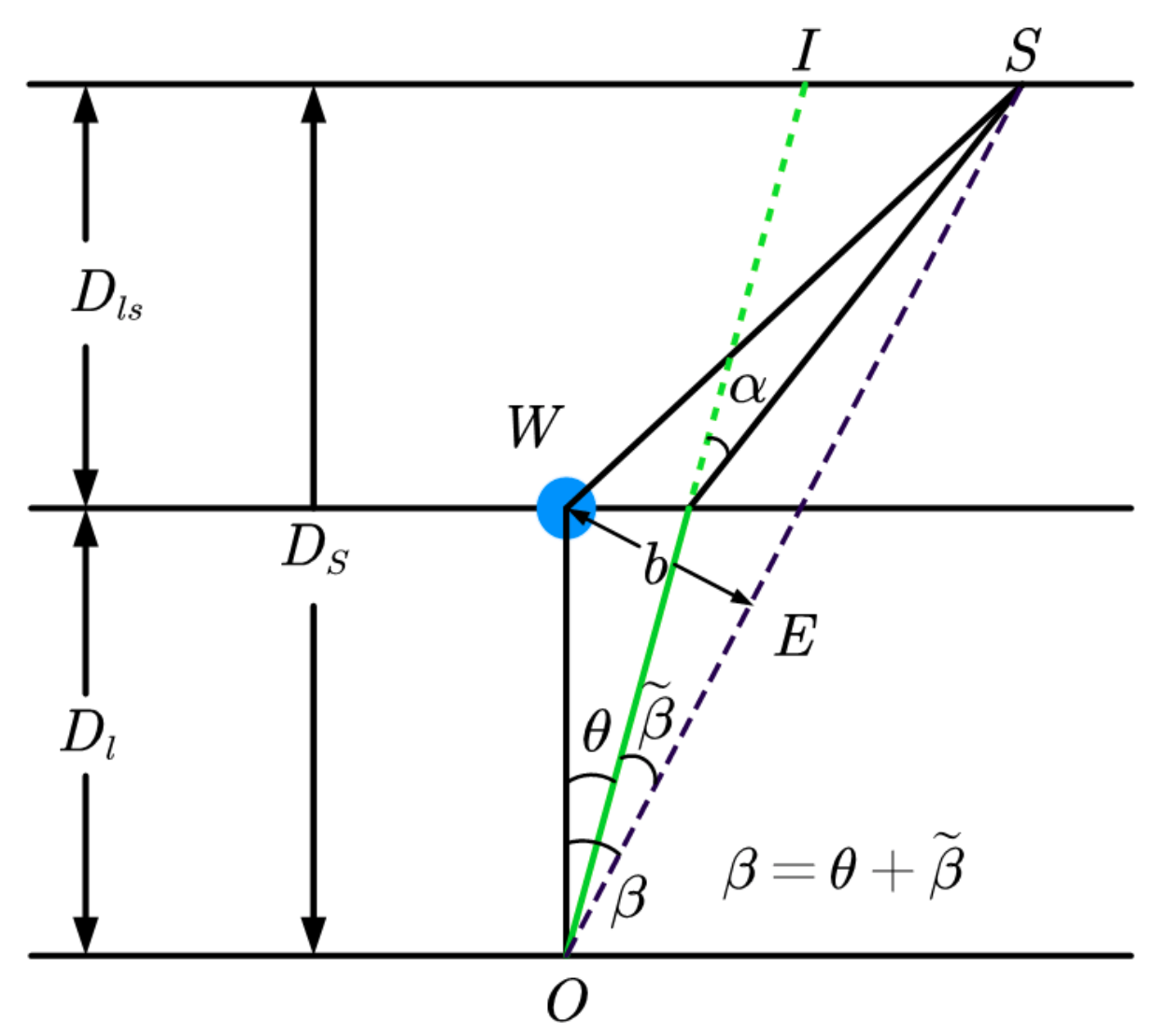}
    \caption{The geometry of a lensing with a pointlike len.}
    \label{fig:lens}
\end{figure}

We depict the lens geometry in Fig. \ref{fig:lens}. We may treat the lens object (in our case the wormhole $W$) to be a point, as long as the scale of lensing geometry is much larger than the lens object. A light ray is emitted at an angle $\beta$ from the source $S$, but deflected by the lens such that the observer $O$ received it at an angle $\theta$. We use the deflection angle $\alpha$ to measure the deflection of light. In the metric \eqref{eq:SSSmetric}, the deflection angle has the following expression (the formulas are generalized by \cite{Virbhadra:1998dy}, compared to the simple case, say, e.g., \cite{Weinberg:1972kfs}):
\begin{equation}
\label{eq:alphadef}
    \alpha = 2 \int_{r_0}^{\infty} \frac{\sqrt{Q(r)}/R(r) dr}{\sqrt{\frac{R^2(r)}{R^2(r_0)} \frac{P(r_0)}{P(r)} - 1}} - \pi ~,
\end{equation}
where $r_0$ is the distance of the closest approach of the light to the center of the gravitational attraction.

Sometimes it is more convenient to use the impact parameter $b$ instead of $r_0$ to describe the light ray. The impact parameter $b$ is defined as
\begin{equation}
    b = R(r_0)/ \sqrt{P(r_0)} ~.
\end{equation}

With the help of \eqref{eq:alphadef}, we can relate the angle $\theta$ and $\beta$ by the lensing geometry. In the weak-field limit we have
\begin{equation}
    D_{ls} \alpha = D_s (\beta - \theta) ~,
\end{equation}
which gives a function $\beta(\theta)$. Observables may then be evaluated from the function $\beta(\theta)$. For example, the magnification is
determined by the ratio between the solid angles
\begin{equation}
    |\mu| = \frac{d\Omega_O}{d\Omega_S} = \Big| \frac{\beta}{\theta} \frac{d\beta}{d\theta} \Big|^{-1}.
\end{equation}

We see that the deflection angle $\alpha$ is essential in the lensing physics. In the following, we shall restrict ourselves to the property of $\alpha$. Moreover, we are interested in the weak-field regime, where we require $b$ to be much larger than any wormhole parameters (in our case $b \gg m$ and $b \gg q$). Thus, we expect the deflection angle to be a series of $m/b$ and $q/b$. In the current paper, we will evaluate $\alpha$ up to the $1/b^4$ order.

\subsection{Deflection angle for special cases}
As pointed out in Sec. \ref{sec:cases}, the asymmetric EBWH reduces to the Schwarzchild black hole and symmetric EBWH in the limits $q = 0$ and $m = 0$, respectively. Since the lensing physics of the latter two objects are thoroughly studied, it would be important to compare our result in the two limiting cases to the previous study.

For our purpose, we will take the massless EBWH as a tool to illustrate different approximation methods in Sec. \ref{sec:massless}. So we shall simply present the deflection angle for a Schwarzschild black hole in the weak-field limit \cite{Frittelli:1999yf,Virbhadra:1999nm}
\begin{align}
    \alpha & \nonumber = \frac{4m}{r_0} + \left( \frac{15\pi}{4} - 4 \right) \left( \frac{m}{r_0} \right)^2 + \left( \frac{122}{3} - \frac{15}{2}\pi \right) \left( \frac{m}{r_0} \right)^3 \\
    & +  \left( \frac{3465}{64} \pi - 130 \right) \left( \frac{m}{r_0} \right)^4 + \mathcal{O} \left( \frac{m}{r_0} \right)^5 ~,
\end{align}
and in terms of the impact parameter $b$, we have
\begin{equation}
\label{eq:alphaSch}
    \alpha =  \frac{4m}{b} + \frac{15\pi}{4} \frac{m^2}{b^2} + \frac{128}{3}  \frac{m^3}{b^3} + \frac{3465}{64} \pi \frac{m^4}{b^4} ~,
\end{equation}
where $m$ is the mass of the Schwarzschild black hole.

The deflection angle of a massless EBWH can be found in sectionSec. \ref{sec:massless}, Eq. \eqref{eq:alpha2ana}.

\section{Massless EBWH as an illustration for different formalisms}
\label{sec:massless}
This section is a pedagogical introduction for the approximation methods, with the massless EBWH metric
\begin{equation}
\label{eq:masslessmetric}
    ds^2 = -dt^2 + dr^2 + (r^2 + q^2) d\Omega_2^2 ~,
\end{equation}
where $q$ is the throat radius and $d\Omega_2^2$ stands for the two-dimensional unit sphere. The gravitational lensing effect of massless EBWH is extensively studied in the literature (see, e.g., \cite{Abe:2010ap,Toki:2011zu,Nakajima:2012pu,Yoo:2013cia,Bozza:2015haa,Lukmanova:2016czn,Tsukamoto:2016qro,Jusufi:2017gyu,Asada:2017vxl,Tsukamoto:2017hva,Bhattacharya:2019kkb}). The deflection angle for the metric \eqref{eq:masslessmetric} is an elliptic function:
\begin{equation}
\label{eq:alphamasslessfull}
    \alpha = 2K\left( \frac{q}{b} \right) - \pi ~,
\end{equation}
where $K(u)$ is the elliptic integral of the first kind, defined as
\begin{equation}
    K(u) \equiv \int_0^1 \frac{dx}{\sqrt{(1-x^2)(1-u^2x^2)}} ~.
\end{equation}

We can write the deflection angle \eqref{eq:alphamasslessfull} in the form
\begin{equation}
    \alpha = \pi \sum_{n=1}^{\infty} \left[ \frac{(2n-1)!!}{(2n)!!} \right]^2 \left( \frac{q}{b} \right)^{2n} ~, 
\end{equation}
where $!!$ denotes the double factorial. Since we are interested in the weak-field region $q \ll b$, we keep the first two contributions
\begin{equation}
\label{eq:alpha2ana}
    \alpha = \frac{\pi}{4} \left( \frac{q}{b} \right)^{2} + \frac{9\pi}{64} \left( \frac{q}{b} \right)^{4} + \mathcal{O} \left( \frac{q}{b} \right)^{6} ~,
\end{equation}

We shall keep the result \eqref{eq:alpha2ana} as fiducial. In the following sections, we shall evaluate the deflection angle of the massless EBWH \eqref{eq:masslessmetric} with the three different formalisms, and compare it with \eqref{eq:alpha2ana}.

\subsection{The AD formalism}
The AD formalism \cite{Amore:2006pi} converts the complicated integrals into a rapidly convergent series of solvable integrals. 

Starting from \eqref{eq:alphadef}, we make a change of variable $z \equiv r_0/r$, and an auxiliary function 
\begin{align}
    V(z) & \nonumber \equiv \frac{z^4}{r_0^2} \frac{R^2(r_0/z)}{Q(r_0/z)} + \frac{P(r_0)r_0^2}{R^2(r_0)} \\
    & - \frac{z^4}{r_0^2} \frac{R^4(r_0/z) P(r_0)}{{Q(r_0/z) P(r_0/z) R^2(r_0)}} ~,
\end{align}
and the expression for $\alpha$ becomes
\begin{equation}
\label{eq:ADintegral}
    \alpha = \int_0^1 \frac{dz}{\sqrt{V(1) - V(z)}} - \pi ~.
\end{equation}

Up to now, the formalism is accurate. Now we write the function $V(z)$ as
\begin{equation}
    V_{\delta}(z) = V_0(z) + \delta (V(z) - V_0(z)) ~,~ \delta \in (0,1) ~.
\end{equation}
where $V_0(z)$ is some function that makes \eqref{eq:ADintegral} solvable (for example $V_0 = \lambda z^2$). If $V(z)$ is of the form $V(z) = \sum_{n=1}^{\infty} v_n z^n$, we can expand the integral as a series of $\delta$, and integrate it term by term. The result, after taking $\delta = 1$, is
\begin{equation}
    \alpha = \frac{3\pi}{2\sqrt{\lambda}} - \frac{1}{\lambda^{3/2}} \sum_{n=1}^{\infty} v_n z^n - \pi ~.
\end{equation}

Finally, we shall assume that the $\alpha$ get the most accurate values when $d\alpha /d\lambda = 0$, i.e. the principle of minimal sensitivity (PMS). The final result is 
\begin{equation}
\label{eq:ADalpha}
    \alpha = \pi \left[ \sqrt{\frac{\pi}{2 \sum_{n=1}^{\infty} v_n z^n}} - 1 \right] ~.
\end{equation}

We elaborate the procedure with the massless EBWH (one may refer to \cite{Dey:2008kn} for more details). Here $-g_{00} = g_{11} = 1$ and $g_{22} = r^2 + q^2$,so the deflection angle is
\begin{equation}
    \alpha = 2 \int_{r_0}^{\infty} \frac{\sqrt{r_0^2 + q^2}dr}{\sqrt{r^2 + q^2}\sqrt{r^2 - r_0^2}} - \pi ~,
\end{equation}
and the function $V(z)$ becomes
\begin{align}
    V(z) = \frac{b^2 - 2q^2}{b^2} z^2 + \frac{q^2}{b^2} z^4 ~.
\end{align}

From the expression \eqref{eq:ADalpha}, we can directly write 
\begin{equation}
    \alpha = \pi \left[ \sqrt{\frac{2b^2}{2b^2 - q^2}} - 1 \right] ~.
\end{equation}
In weak-field approximation $q \ll b$, the final result is
\begin{equation}
\label{eq:ADmassless}
    \alpha = \frac{\pi}{4} \left( \frac{q}{b} \right)^2 + \frac{3\pi}{32} \left( \frac{q}{b} \right)^4 + \mathcal{O} \left( \frac{q}{b} \right)^6 ~.
\end{equation}

The result is accurate in the first order, but disagrees with \eqref{eq:alpha2ana}. The difference is noticed in \cite{Nakajima:2012pu}, and explained as the failure of PMS near the throat $r=0$.

\subsection{The GBT formalism}
\begin{figure}[ht]
    \centering
    \includegraphics[width=\linewidth]{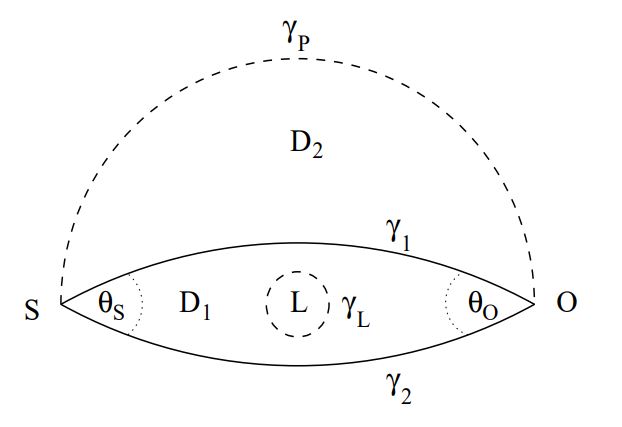}
    \caption{The lensing geometry in weak-field approximation. The two geodesics $\gamma_1$ and $\gamma_2$ represent light rays from the source $S$ to the observer. The domain $D_1$ contains the lens $L$ while $L \notin D_2$. The two domains intersect at $\gamma_1$, and we require $D_2$ to be asymptotically flat at least in the neighbor of $S$ and $O$. Finally, we need an auxiliary curve $\gamma_L$ enclosing $L$, and $\gamma_P$, which is the boundary curve of $D_2$.}
    \label{fig:gbtgeo}
\end{figure}

The GBT formalism has been widely applied to lensing physics since it manifests the physics in a topological viewpoint. Here we show the lensing geometry in Fig. \ref{fig:gbtgeo} from \cite{Gibbons:2008rj}.

The strategy of GBT formalism is as follows. First, the Gauss-Bonnet theorem for a domain $D$ is
\begin{equation}
\label{eq:GBTformal}
    \iint_D K dS + \int_{\partial D} \kappa dt + \sum_i \phi_i = 2\pi \chi(D) ~.
\end{equation}
Here, $K$ and $\kappa$ are Gaussian curvature and geodesic curvature. The angle $\alpha_i$ are the exterior angles, while $\chi(D)$ is the Euler characteristic number. The line integral is done with respect to the affine parameter $t$ (not to be confused with the $t$ coordinate). The exterior angle is related to the interior angle by $\theta_S = \pi - \phi_S$ and $\theta_O = \pi - \phi_O$.

For our case, we shall assume the lens is nonsingular so that $\chi (D)=1$. Notice that the geodesic curvature for a geodesic vanishes, so that $\kappa(\gamma_1) = \kappa(\gamma_2) = 0$. The formula \eqref{eq:GBTformal} for domain $D_1$ then gives
\begin{equation}
    \theta_S + \theta_O = \iint_{D_1} K dS ~,
\end{equation}
which may help us to intuitively judge whether the lens is convex or concave.

Applying the formula \eqref{eq:GBTformal} to the domain $D_2$, we choose sufficiently remote $S$ and $O$ such that $\theta_S \simeq \theta_O \simeq \pi/2$, so that
\begin{equation}
    \int_{\gamma_P} \kappa (\gamma_P) dt = \pi - \iint_{D_2} KdS ~.
\end{equation}
Finally, for asymptotically flat domain $D_2$, we have $\kappa (\gamma_P) dt/ d\alpha = 1$, so the line integral is
\begin{equation}
    \int_{\gamma_P} \kappa (\gamma_P) dt = \int_{\gamma_P} d\alpha = \pi + \alpha ~,
\end{equation}
so the deflection angle can be directly obtained by
\begin{equation}
\label{eq:GBTa}
    \alpha = - \iint_{D_2} KdS ~.
\end{equation}

Before proceeding, we shall mention that, the relation between the deflection angle $\alpha$ from GBT formalism \eqref{eq:GBTa} and the conventional one \eqref{eq:alphadef} is not clear at first glance. Fortunately, there are vast literatures on the topic, where the equivalence of the two deflection angle is established in generic spacetime \cite{Ishihara:2016vdc,Ishihara:2016pi, Ishihara:2016sfv,Ono:2017pi,Ono:2018ybw,Ono:2018jrv,Ono:2019pi, Takizawa:2020pi,Takizawa:2020dja,Takizawa:2020egm}. Therefore, in the EBWH case, we can simply take the deflection angle evaluated by \eqref{eq:GBTa} to be identical to the conventional one \eqref{eq:alphadef}.

It is generically convenient to evaluate the surface integral in the optical metric. We elaborate the process by the massless EBWH. For convenience we fix $\theta = \pi/2$ due to the spherical symmetry, so the null geodesic $ds^2 = 0$ gives 
\begin{equation}
    dt^2 = dr^2 + (r^2 + q^2) d\varphi^2 ~.
\end{equation}
The next step is to write the null geodesics in the form
\begin{equation}
    dt^2 = h_{ab} d\lambda^a d\lambda^b = du^2 + \zeta^2(u) d\varphi^2 ~,
\end{equation}
and for massless EBWH, we already have $u = r$, $\zeta(u) = \sqrt{u^2 + q^2}$, so the Gaussian optical curvature is
\begin{align}
\label{eq:Kmassless}
    K & \nonumber \equiv - \frac{1}{\zeta(u)} \left[ \frac{dr}{du} \frac{d}{dr} \left( \frac{dr}{du} \right) \frac{d\zeta}{dr} + \left( \frac{dr}{du} \right)^2 \frac{d^2 \zeta}{dr^2} \right] \\
    & = -\frac{1}{\zeta} \frac{d^2 \zeta}{dr^2} = - \frac{q^2}{(q^2 + r^2)^2} ~,
\end{align}
and the deflection angle is
\begin{equation}
   \alpha \equiv -\iint_{D_2} KdS = -\int^\pi_0 d\varphi\int^\infty_{r_{\text{OE}}}K\sqrt{h}dr  ~.
\end{equation}
It is common to use straight line approximation, i.e. $r_{\text{OE}}=b/\sin\varphi$, which is precise at leading order. Unfortunately, the straight line approximation will cause considerable deviation at higher order \cite{Jusufi:2017gyu}. Since we are interested in the higher-order correction to the deflection angle, we need to introduce $u=1/r$ and expand it into higher-order series of $1/b$ \cite{Ono:2017pie} in our calculation. After some straightforward calculations, we have
\begin{equation}
    u=\frac{1}{r}=\frac{\sin\varphi}{b}+\Big[\big(\frac{\pi}{2}-\varphi\big)\cos\varphi+\sin\varphi\Big]\frac{q^2}{b^3} ~.
\end{equation}
Now we can calculate the deflection angle
\begin{align}
\label{eq:alphagbtm=0}
   \alpha & \nonumber =-\int^\pi_0 d\varphi\int^{\frac{\sin\varphi}{b}+[(\frac{\pi}{2}-\varphi)\cos\varphi+\sin\varphi]\frac{q^2}{b^3}}_0 K\sqrt{h}u^2 du \\
     &  \simeq  \frac{\pi}{4} \left( \frac{q}{b} \right)^2 + \frac{39\pi}{64} \left( \frac{q}{b} \right)^4 + \mathcal{O} \left( \frac{q}{b} \right)^6 ~.
\end{align}
This has to be compared by the result from straight line approximation \cite{Jusufi:2017gyu}: 
\begin{equation}
	\alpha = \frac{\pi}{4} \left( \frac{q}{b} \right)^2 - \frac{9\pi}{64} \left( \frac{q}{b} \right)^4 + \mathcal{O} \left( \frac{q}{b} \right)^6 ~,
\end{equation}	
and we see that the improved result \eqref{eq:alphagbtm=0} also deviates from the precise result in massless case \eqref{eq:alpha2ana}. However, in later sections we will show that,  the improved GBT formalism in isotropic coordinates will result in an accurate result up to $q^4/b^4$ in massless limits.

\subsection{The PPN formalism}

The PPN formalism is based on a simple observation that, in the weak-field limit, a wild range of physical quantities can be expressed as a series of the effective Newtonian potential. For a gravitational lens with mass $M$, the effective Newtonian potential is $\Phi = M/r$. Notice that we have used Planck units with $8\pi G = 1$, so in principle, the coefficient $G$ in the definition of $\Phi$ should lead to a factor of $1/8\pi$. However, since $1/8\pi$ is only a constant, we may suppress it for convenience by a rescaling. This is equivalent to transfer the $1/8\pi$ factor from $\Phi$ into the coefficients of $\Phi$ in the PPN series.

Before proceeding, we shall clarify some tricky issues. First, as we shall see from \eqref{eq:Mpm}, the gravitational masses observed in Universes I and II are different. Thus we shall specify which universe we are working with before proceeding. Besides, for a complicated metric such as EBWH, the expression for $\Phi$ is not obvious. For example, the physical radius of  EBWH is $R$ instead of the $r$ from \eqref{eq:ebsol}, so it is not clear whether we should use $r$ or $R$ in the definition of $\Phi$. Besides, the form of mass could change in different conventions. For example, in some studies, e.g., \cite{Nandi:2016uzg}, the convention for EBWH is chosen such that the Kepler mass in one side is $m$ and in the other side is $-m e^{\pi m/\sqrt{{q^2 - m^2}}}$. Thus if we apply different conventions, the definition for $\Phi$ seems to be varying.

In view of the above argument, we need to be more careful about what is an effective Newtonian potential. In the standard treatment of PPN formalism, we start with the isotropic metric
\begin{equation}
\label{eq:isodef}
    ds^2 = -A(l)dt^2 + B(l) \left( dl^2 + l^2 d\Omega_2^2 \right) ~.
\end{equation}
The isotropic coordinate is defined such that the spatial part of the metric is conformally flat, so that the light cone appears round. In the rest of this paper, we will use $l$ to represent the radial coordinate as long as we are working with isotropic coordinates. 

Let us illustrate how the PPN formalism works in the massless case. The metric for massless EBWH in isotropic coordinate is (see appendix \ref{app:isotropic} for more details)

\begin{equation}
   ds^2 = -dt^2 + \Big(1+\frac{q^2}{4l^2}\Big)^2 \left( dl^2 + l^2 d\Omega_2^2 \right) ~.
\end{equation}

As one can see, we can simply define the effective Newtonian potential $\Phi \propto q^2/l^2$. Notice that $\Phi$ shows a $q^2/l^2$ dependence, instead of the conventional $q/l$ one. This is expected since the deflection angle \eqref{eq:alpha2ana} contains only $q^2/b^2$ term. Let's naively take $\Phi = q^2/4 l^2$, and the metric coefficients become
\begin{equation}
     A(\Phi) = 1 ~,~ B(\Phi) =1 + 2\Phi + \Phi^2 ~.
\end{equation}

Under the isotropic metric above, the deflection angle should be expressed as
\begin{equation}
\alpha = 2 \int_{r_0}^{\infty}\frac{1}{r^2} \sqrt{\frac{A B}{1/b^2 - A/r^2}} dr - \pi~.
\end{equation}
Finally, only the nonzero even order terms remain. So the deflection angle in PPN formalism is
\begin{align}
\alpha = \frac{\pi}{4}\left(\frac{q}{b}\right)^2 + \frac{9}{64} \pi  \left(\frac{q}{b}\right)^4 + \mathcal{O}\left(\frac{q}{b}\right)^6~.  
\end{align}
Compared to \eqref{eq:alpha2ana}, the result in PPN formalism is relatively accurate considering the massless case.

From the above argument, it seems that we need two PPN parameters to describe the massive EBWH, since it contains two model parameters, $q$ and $m$. Fortunately, as we shall see in Sec. \ref{sec:PPN}, it is still possible to use one single PPN parameter to describe the lensing physics in one side of EBWH.

\section{Deflection angle of massive EBWH}
\label{sec:massive}
Now we come to the generic EBWH. Our strategy is as follows. We use the AD formalism, the GBT formalism, and the PPN formalism to evaluate the deflection angle up to second order, in Sec. \ref{sec:AD}, \ref{sec:GBT} and \ref{sec:PPN}, respectively. Then, we numerically evaluate the deflection angle, and compare the numerical result to the above methods in Sec. \ref{sec:num}.

Before proceeding, we mention that the metric component in the isotropic coordinate is (see Appendix \ref{app:isotropic} for details)
\begin{equation}
\label{eq:Aiso}
    A = \exp \left[ \gamma \left(\pi - 4 \arctan\frac{l\gamma}{m} \right) \right] ~,
\end{equation}
\begin{equation}
    B = \frac{1}{A} \frac{\left(l^2 \gamma^2 + m^2 \right)^2}{4l^4\gamma^4} ~.
\end{equation}
with the dimensionless constant being
\begin{equation}
    \gamma \equiv \frac{m}{\sqrt{q^2 - m^2}} ~.
\end{equation}

For the AD and GBT formalism, it would be useful to do the calculation in the two coordinate system as a consistency check.

Finally, when presenting the final result, we will meet the following combination:
\begin{equation}
\label{termform}
    (\pm 1)^{\alpha+\beta} \frac{m^{\alpha} q^{\beta}}{b^{\alpha + \beta}} e^{(\pm 1)^{\alpha + \beta} \frac{m\pi}{\sqrt{q^2 - m^2}}} ~.
\end{equation}
Such a combination seems more complicated than expected, and that is because the metric coefficients on each side of the wormhole cannot be simultaneously set to asymptotic unity. For simplicity we define the rescaled impact parameter as
\begin{equation}
    b_{\pm} \equiv \pm b e^{\pm \frac{m\pi}{\sqrt{q^2 - m^2}}} ~,
\end{equation}
where $\pm$ correspond to wormhole sides with positive/negative mass as mentioned above.
Now the expression \eqref{termform} just simplifies to
\begin{equation}
\label{eq:finalform}
    \frac{m^{\alpha} q^{\beta}}{b_{\pm}^{\alpha + \beta}}.
\end{equation}

Compared to the massless case, the mass is introduced and the impact parameter is rescaled in massive results.

\subsection{The AD formalism}
\label{sec:AD}

The auxiliary function $V(z)$ in the metric \eqref{eq:ebsol} is
\begin{align}
\label{eq:Vpmz}
    V_{\pm}(z) & \nonumber = \left[ r_0^2 + z^2(q^2 - m^2) \right] \frac{z^2}{r_0^2} \Bigg\{ 1 - \frac{q^2 - m^2 + r_0^2/z^2}{r_0^2 + q^2 - m^2} \\
    & \times  \exp \left[ \pm 4\gamma \left( \arctan \frac{r_0 \gamma}{m} - \arctan \frac{r_0 \gamma}{mz} \right) \right] \Bigg\} ~.
\end{align}
We are only interested in the deflection angle up to the $1/b^4$ term, so we expand \eqref{eq:Vpmz} up to $z^4$ order, substitute into it into formula \eqref{eq:ADalpha}, we get
\begin{align}
    & \ \ \ \ \alpha  \nonumber \simeq \pm \frac{4m}{r_0} + \frac{\pi^2}{4} \frac{q^2}{r_0^2} + \left( \frac{24}{\pi} - 16 + \frac{15}{4} \pi \right) \frac{m^2}{r_0^2} \\
    & \nonumber \pm \left( \frac{37}{3} - 4\pi \right) \frac{q^2m}{r_0^3} \pm \left( \frac{160}{\pi^2} - \frac{192}{\pi} + 89 - 12\pi \right) \frac{m^3}{r_0^3} \\
    & \nonumber - \frac{5\pi}{32} \frac{q^4}{r_0^4} + \left( \frac{142}{\pi} - 92 + \frac{245}{16} \pi \right) \frac{q^2m^2}{r_0^4} \\
    & + \left( \frac{1120}{\pi^3} - \frac{1920}{\pi^2} + \frac{1362}{\pi} - 420 + \frac{1563}{32} \pi \right) \frac{m^4}{r_0^4} ~,
\end{align}
and translate into the rescaled impact parameter $b_\pm$, where we have
\begin{align}
\label{eq:alphaADr}
    \alpha & \nonumber = 4 \frac{m}{b_{\pm}} + \frac{\pi}{4} \frac{q^2}{b_{\pm}^2} + \left( \frac{24}{\pi} - 8 + \frac{15}{4}\pi \right) \frac{m^2}{b_{\pm}^2} \\
    & \nonumber + \left( \frac{160}{\pi^2} - \frac{96}{\pi} + 47 + 3\pi \right) \frac{m^3}{b_{\pm}^3} + \left( \frac{43}{3} - 3\pi \right) \frac{m}{b_{\pm}} \frac{q^2}{b_{\pm}^2} \\
    & \nonumber + \left( \frac{1120}{\pi^3} - \frac{960}{\pi^2} + \frac{570}{\pi} - 54 + \frac{1059}{32} \pi \right) \frac{m^4}{b_{\pm}^4} \\
    & + \left( \frac{166}{\pi} - \frac{62}{3} - \frac{19}{16} \pi \right)\frac{m^2}{b_{\pm}^2} \frac{q^2}{b_{\pm}^2} + \frac{3\pi}{32} \frac{q^4}{b_{\pm}^4} ~.
\end{align}

One may also use the formalism in the isotropic coordinate by a similar procedure and get
\begin{align}
\label{eq:alphaADl}
    \alpha & \nonumber = 4 \frac{m}{b_{\pm}} + \frac{\pi}{4} \frac{q^2}{b_{\pm}^2} + \left( \frac{24}{\pi} - 8 + \frac{15}{4}\pi \right) \frac{m^2}{b_{\pm}^2} \\
    & \nonumber + \left( \frac{160}{\pi^2} - \frac{96}{\pi} + 56 \right) \frac{m^3}{b_{\pm}^3} + \frac{16}{3} \frac{m}{b_{\pm}} \frac{q^2}{b_{\pm}^2} \\
    & \nonumber + \left( \frac{1120}{\pi^3} - \frac{960}{\pi^2} + \frac{678}{\pi} - 72 + \frac{1737}{64} \pi \right) \frac{m^4}{b_{\pm}^4} \\
    & + \left( \frac{58}{\pi} - \frac{8}{3} + \frac{151}{32} \pi \right)\frac{m^2}{b_{\pm}^2} \frac{q^2}{b_{\pm}^2} + \frac{9\pi}{64} \frac{q^4}{b_{\pm}^4} ~.
\end{align}

In both cases, the results \eqref{eq:alphaADr} and \eqref{eq:alphaADl} agree with the previous result from AD formalism \eqref{eq:ADmassless} in the massless limit. Compareing the results \eqref{eq:alphaADr} and \eqref{eq:alphaADl} to the precise one: in the $q=0$ limit, they differ from \eqref{eq:alphaSch} at the order $m^2/b^2_\pm$; in the $m=0$ limit, they differ from \eqref{eq:ADmassless} at the $q^4/b^4_\pm$ order. Moreover, the AD formalism gives different results in different coordinate systems at the $1/b^3_\pm$ order, even if the underlying metrics are the same.

\subsection{The GBT formalism}
\label{sec:GBT}

The Gaussian curvature \cite{Ovgun:2018fnk} and orbit equation are
\begin{equation}
    K = \frac{m(m \mp 2r)-q^2}{(r^2 + q^2 - m^2)^2}\exp\left({\pm4\gamma\arctan{\frac{\gamma r}{m}}}\right) ~,
\end{equation}
\begin{align}
   u& \nonumber =\frac{\sin\varphi}{be^{\pm\pi\gamma}}\pm\frac{2m}{b^2e^{2\pm\pi\gamma}}+\Big[(q^2+3m^2)\big(\frac{\pi}{2}-\varphi\big)\cos\varphi\\
    & \nonumber +(q^2+5m^2)\sin\varphi\Big]\frac{1}{b^3e^{3\pm\pi\gamma}}\pm\big[9(q^2+7m^2)\\ 
    & -(q^2-9m^2)\cos2\varphi\big]\frac{m}{3b^4e^{4\pm\pi\gamma}} ~,
\end{align}
so the integration directly gives
\begin{align}
\label{eq:alphagbtr}
   \alpha & \nonumber = \frac{4m}{b_{\pm}} + \frac{\pi}{4} \frac{q^2 +15 m^2}{b_{\pm}^2} + \frac{8m}{3b_{\pm}} \frac{4q^2 +11m^2}{b_{\pm}^2} \\
    & + \frac{3\pi}{64} \frac{(13q^4+38q^2m^2-51m^4 )}{b_{\pm}^4} ~.
\end{align}

When $m = 0$, we recover the result in \eqref{eq:alphagbtm=0}. Also, when $q=0$, the result differs from \eqref{eq:alphaSch} even at the $m^2/b^2_\pm$ order.

We may repeat the procedure in the isotropic coordinate, where the Gaussian curvature and orbit equation are
\begin{align}
 K =&\nonumber \frac{128l^3[q^2(\pm m+2l)+m(\mp m^2+2ml+4l^2)] }{\left( q^2 - m^2 + 4l^2  \right)^4}\\& \times \exp\left[{\mp2\gamma\left(\pi-4\arctan\frac{2\gamma l}{m}\right)} \right] ~,
\end{align}
\begin{align}
  u& \nonumber =\frac{\sin\varphi}{be^{\pm\pi\gamma}}\pm\frac{2m}{b^2e^{2\pm\pi\gamma}}+\Big[(q^2+15m^2)\big(\frac{\pi}{2}-\varphi\big)\cos\varphi\\
    & \nonumber +(q^2+23m^2)\sin\varphi\Big]\frac{1}{4b^3e^{3\pm\pi\gamma}}\pm\big[9(3q^2+29m^2)\\ 
    & +(5q^2+27m^2)\cos2\varphi\big]\frac{m}{12b^4e^{4\pm\pi\gamma}} ~.
\end{align}
After evaluating the integral and substituting $l$ into $r$, we get
\begin{align}
\label{eq:alphagbtl}
   \alpha & \nonumber = \frac{4m}{b_{\pm}} + \frac{\pi}{4} \frac{q^2 +15 m^2}{b_{\pm}^2} + \frac{8m}{3b_{\pm}} \frac{2q^2 +13 m^2}{b_{\pm}^2} \\
    & - \frac{3\pi}{64} \frac{q^4+186q^2m^2 +835 m^4}{b_{\pm}^4} ~.
\end{align}
As we shall see, the results \eqref{eq:alphagbtr} and \eqref{eq:alphagbtl} differ at the order $1/b^3_\pm$.

Before proceeding, let's briefly comment on the result. From \eqref{eq:alpha2ana} we see that, in the massless limit $m=0$, the accurate coefficient of the $q^4/b^4$ term is $9\pi /64$, while the GBT method in conventional coordinate and isotropic coordinate yield $39\pi/64$ and $-3\pi/64$, respectively. Therefore, we can intuitively deduce that the GBT formalism is more accurate in the isotropic coordinate.

\subsection{The PPN formalism}
\label{sec:PPN}
To proceed with the PPN formalism we need to know the Newtonian potential $\Phi$. However, it appears tricky to define $\Phi$ in our case. Recall that the throat is located at $r=-m$, i.e., $l = (q-m)/2$; and the two asymptotic region $r \to \pm \infty$ corresponds to $l = 0$ and $l \to \infty$, respectively. Thus, simple expressions such as $\Phi = m/2l$ would not work for both sides. In fact, by the conventional treatment, we shall define a Newtonian potential that vanishes at a flat region. However, we cannot set the two sides of the universe in \eqref{eq:ebsol} to be asymptotically flat simultaneously. Therefore, we cannot expect a global Newtonian potential satisfying both $\Phi(l=0) = 0$ and $\Phi(l\to\infty) = 0$.

Therefore, we shall use an alternative approach. We shall define the two Newtonian potential, $\Phi_-$ corresponds to the $r>-m$ universe with negative mass, and $\Phi_+$ corresponds to the $r<-m$ universe with positive mass. Moreover, as we see in Appendix \ref{app:isotropic}, Universe II can be transferred to Universe I by a change of variable $l \to (q^2-m^2)/4l$. So without loss of generality let us first work out the $\Phi_-$ case.

In the $\Phi_-$ case $l > (q-m)/2$ , and the weak-field limit is valid at the region $l \gg q$. Now for the large $l$, the $\arctan$ function in \eqref{eq:Aiso} approximates to $\pi/2$, so let us rewrite the coefficient $A(l)$ as
\begin{equation}
    A(l) = \exp \left[ \gamma \left( 4 \arctan \frac{m}{2\gamma l} - \pi \right) \right] ~.
\end{equation}
Notice that at the asymptotic region $A(\infty) = \exp (-\gamma \pi) \neq 1$, this is the cost we have to pay for the metric \eqref{eq:ebsol}, where the two sides are written as symmetric as possible. Now, one may simply define 
\begin{equation}
    \Phi_- = -\frac{m}{2\gamma l} ~,~ l \in (\frac{q-m}{2},\infty) ~,
\end{equation}
and we may expand the metric components as
\begin{align}
A(\Phi) \nonumber & = e^{-\gamma \pi} \left[1 - 4 (\gamma \Phi_-) + 8 (\gamma \Phi_-)^2 \right.\\ 
\nonumber & - \frac{1}{3}\left(32 - \frac{4}{\gamma^2}\right)(\gamma \Phi_{-})^3 \\
& + \frac{16}{3} \left(2 - \frac{1}{\gamma^2}\right)(\gamma \Phi_{-})^4 + \mathcal{O}(\gamma \Phi_{-})^5]~.
\end{align}
The deflection angle for this side is then
\begin{align}
\alpha & \nonumber = \frac{4m}{b_-} + \frac{\pi}{4}\left(16 + \frac{1}{\gamma^2}\right)\frac{m^2}{b_-^2} + \frac{16}{3}\left(9 + \frac{1}{\gamma^2}\right)\frac{m^3}{b_-^3}\\
& + \left(64 + \frac{10}{\gamma^2} + \frac{9}{64 \gamma^2}\right)\pi \frac{m^4}{b_-^4} + \mathcal{O}\left(\frac{m^5}{b_-^5}\right)~.
\end{align}
For the positive mass side, it corresponds to $r\to-\infty$ and $l\to 0$. To avoid the counterintuition of representing spatial infinity near coordinate origin we take the transformation $l \to (q^2-m^2)/4l$, and the metric coefficient becomes
\begin{equation}
    A(l)= \exp \left[ \gamma \left(\pi- 4 \arctan \frac{m}{2\gamma l} \right) \right] ~,
\end{equation}
\begin{equation}
   \Phi_+ = \frac{m}{2\gamma l} ~,~ l \in (\frac{q+m}{2},\infty) ~.
\end{equation}

In terms of effective potential $\Phi_+$,
\begin{align}
A(\Phi) \nonumber & = e^{\gamma \pi} \left[1 - 4 (\gamma \Phi_{+}) + 8 (\gamma \Phi_{+})^2 \right.\\ 
\nonumber & - \frac{1}{3}\left(32 - \frac{4}{\gamma^2}\right)(\gamma \Phi_{+})^3 \\
& + \frac{16}{3} \left(2 - \frac{1}{\gamma^2}\right)(\gamma \Phi_{+})^4 + \mathcal{O}(\gamma \Phi_{+})^5]~.
\end{align}
and the deflection angle for this side is
\begin{align}
\alpha & \nonumber = \frac{4m}{b_+} + \frac{\pi}{4}\left(16 + \frac{1}{\gamma^2}\right)\frac{m^2}{b_+^2} + \frac{16}{3}\left(9 + \frac{1}{\gamma^2}\right)\frac{m^3}{b_+^3}\\
    & + \left(64 + \frac{10}{\gamma^2} + \frac{9}{64 \gamma^2}\right)\pi \frac{m^4}{b_+^4} + \mathcal{O}\left(\frac{m^5}{b_+^5}\right)~.
\end{align}
After using the expression of $\gamma$, we finally get
\begin{align}
\label{eq:alphappn}
    \alpha & \nonumber = \frac{4 m}{b_{\pm}} + \frac{15\pi}{4} \frac{m^2}{b_{\pm}^2} + \frac{\pi}{4} \frac{q^2}{b_{\pm}^2} + \frac{128}{3} \frac{m^3}{b_{\pm}^3} + \frac{16}{3} \frac{m}{b_{\pm}} \frac{q^2}{b_{\pm}^2} \\
    & + \frac{3465}{64} \pi \frac{m^4}{b_{\pm}^4} + \frac{311}{32} \pi \frac{m^2}{b_{\pm}^2} \frac{q^2}{b_{\pm}^2} + \frac{9\pi}{64} \frac{q^4}{b_{\pm}^4} ~.
\end{align}
We comment that the result \eqref{eq:alphappn} returns to the precise solutions \eqref{eq:alphaSch} and \eqref{eq:alpha2ana} in the limit $q \to 0$ and $m \to 0$, respectively.

\subsection{Numerical evaluation}
\label{sec:num}
Before proceeding, we shall set a ``fiducial'' result for $\alpha$, by expanding \eqref{eq:alphadef} as a series of $q/b_\pm$ and $m/b_\pm$ by brutal force. The result is simply
\begin{align}
\label{eq:alphaana}
    \alpha & \nonumber = 4 \frac{m}{b_{\pm}} + \frac{15\pi}{4} \frac{m^2}{b_{\pm}^2} + \frac{\pi}{4} \frac{q^2}{b_{\pm}^2}  + \frac{128}{3} \frac{m^3}{b_{\pm}^3} + \frac{16}{3} \frac{m}{b_{\pm}} \frac{q^2}{b_{\pm}^2} \\
    & + \frac{3465}{64} \pi \frac{m^4}{b_{\pm}^4} + \frac{311}{32} \pi \frac{m^2}{b_{\pm}^2} \frac{q^2}{b_{\pm}^2} + \frac{9\pi}{64} \frac{q^4}{b_{\pm}^4} ~.
\end{align}
Not surprisingly, the result \eqref{eq:alphaana} coincides with the PPN result \eqref{eq:alphappn}. In PPN formalism we expand the deflection angle as a series of $\Phi_{\pm}$, and $\Phi_{\pm}$ itself can be seen as a series of $m/b_{\pm}$ and $q/b_{\pm}$. Therefore, the procedure should be equivalent to expand $\alpha$ with respect to $m/b_{\pm}$ and $q/b_{\pm}$ directly, and we then recover \eqref{eq:alphaana}.

Surely, we have no reason to claim \eqref{eq:alphaana} to be precise up to the order $1/b^4_\pm$. However, as we shall show in the following, the result \eqref{eq:alphaana} shows the least deviation on the numerical result, so we may comprehensively understand the error in different methods, by comparing their coefficients with that in \eqref{eq:alphaana}.

\begin{figure}[ht]
    \centering
    \includegraphics[width=0.96\linewidth]{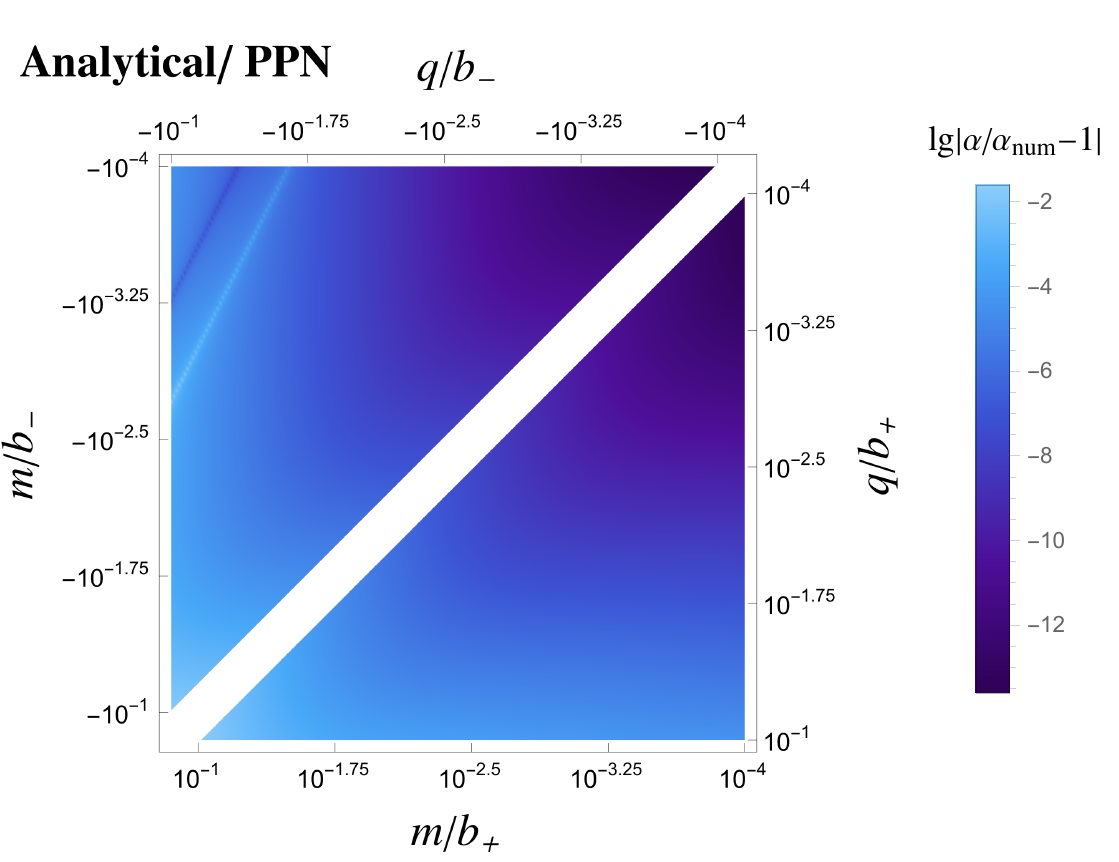}
    \caption{Relative error of analytical/ PPN expansion $\alpha$ compared with numerical result $\alpha_{\text{num}}$ for both sides of EBWH.}
    \label{fig:ana}
\end{figure}

As shown in Fig. \ref{fig:ana}, the approximation of analytical expansion \eqref{eq:alphaana} is satisfying, while the relative error quickly decreases as parameters $m/b_{\pm}$ and $q/b_{\pm}$ decrease. In isotropic coordinates the analytical expansion is identical, and thus Fig. \ref{fig:ana} still fits.

The upper triangle region of Fig. \ref{fig:ana} is the negative mass side, and the lower region represents the positive mass side. On the positive mass side the relative error changes smoothly, while on the other side it does not. There is a light thin line on the negative mass side at which the relative error is larger than neighboring regions, which is because the deflection angle reaches zero and changes its sign when crossing this line, causing the machine error to dominate around this area. There is also a dark line on the negative mass side, and this is simply the area where analytical expansion reaches its best approximation. We will see similar patterns for other approximation methods.

\begin{figure}[ht]
    \centering
    \includegraphics[width=0.96\linewidth]{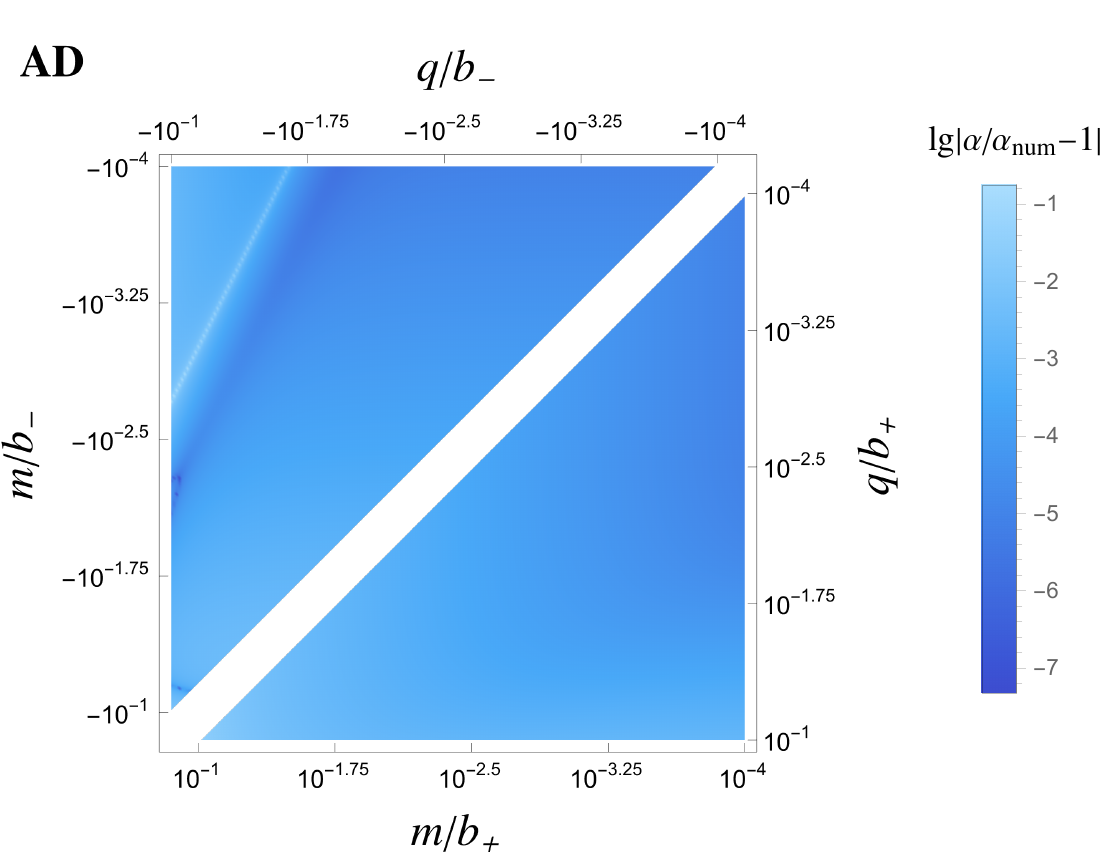}
    \includegraphics[width=0.96\linewidth]{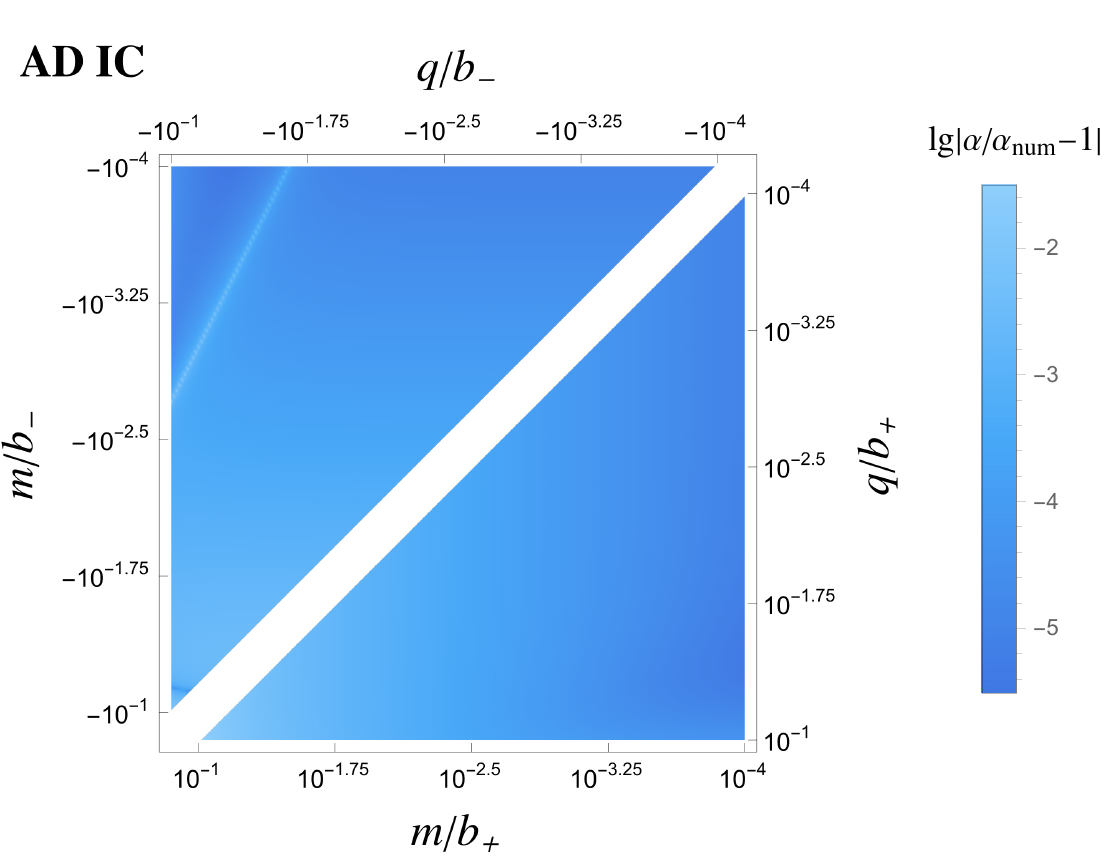}
    \caption{Relative error of AD formalism result compared with numerical result $\alpha_{\text{num}}$ for both sides of EBWH.}
    \label{fig:ad}
\end{figure}

\begin{figure}[ht]
    \centering
    \includegraphics[width=0.96\linewidth]{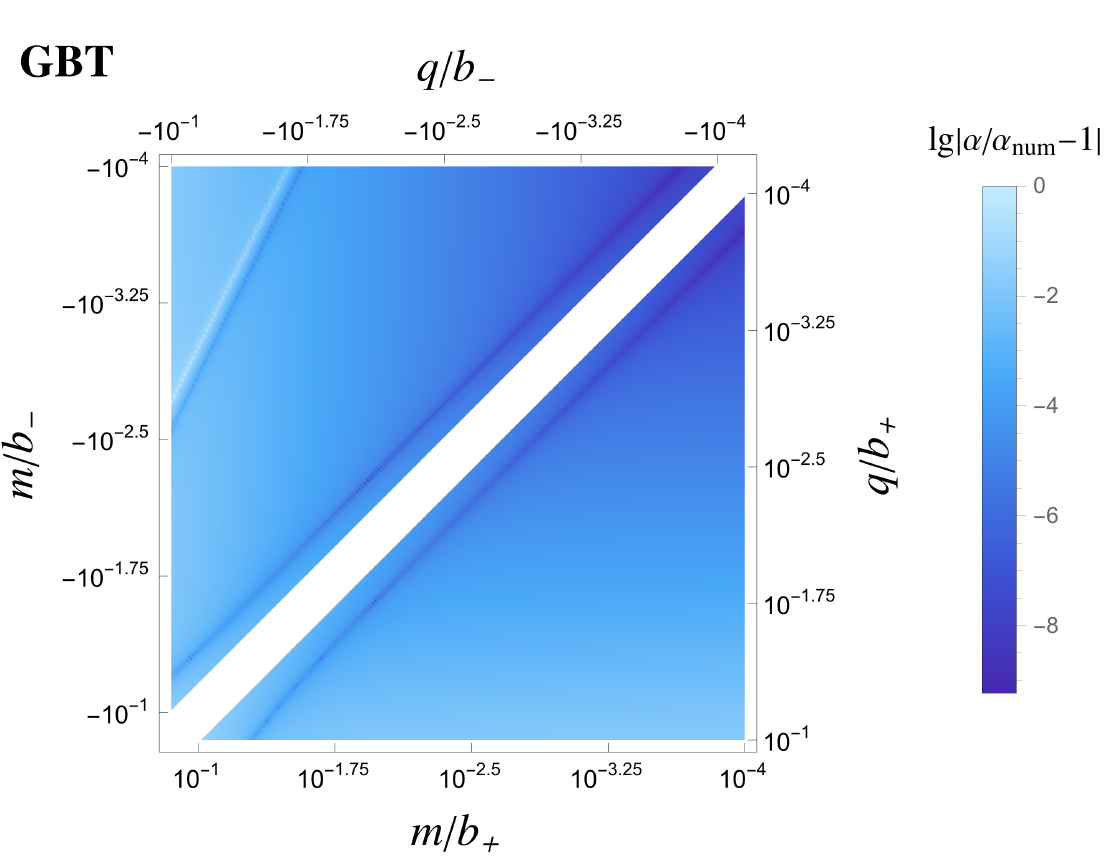}
    \includegraphics[width=0.96\linewidth]{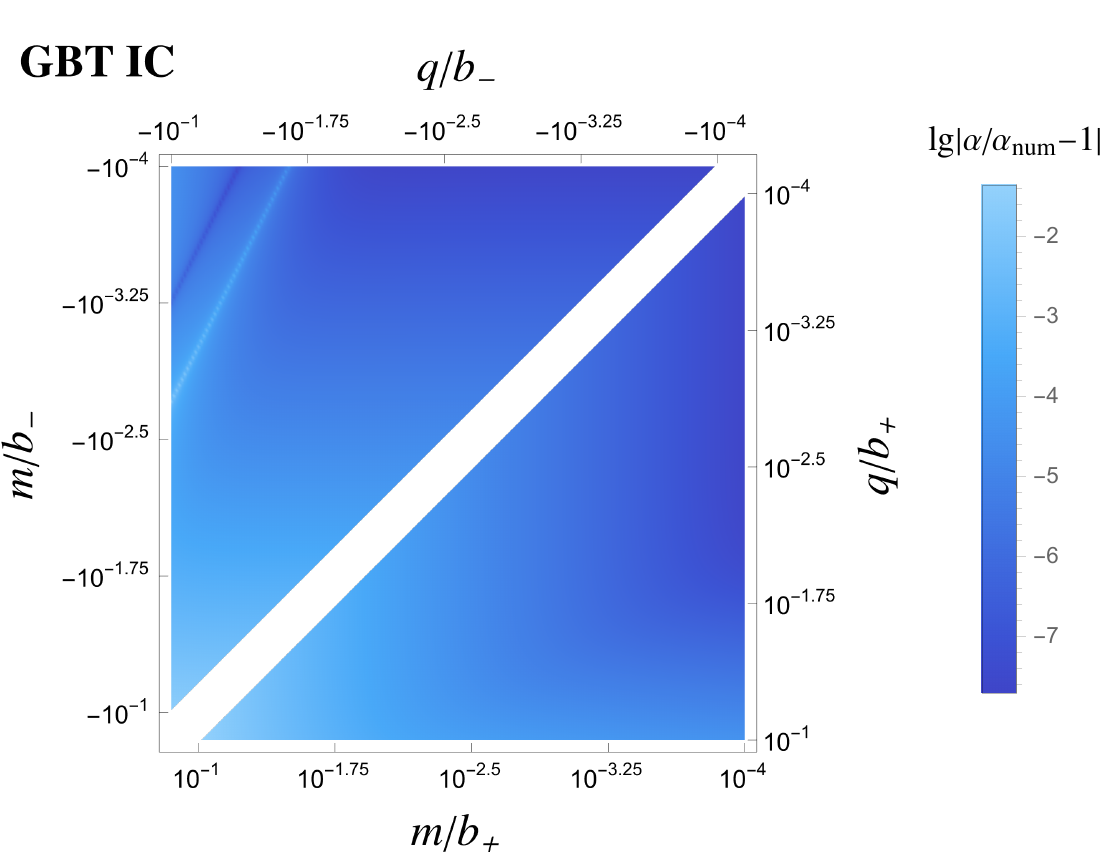}
    \caption{Relative error of GBT formalism result compared with numerical result $\alpha_{\text{num}}$ for both sides of EBWH.}
    \label{fig:gbt}
\end{figure}

We depict the numerical results from AD formalism in Fig. \ref{fig:ad} and GBT formalism in Fig. \ref{fig:gbt}, respectively. The PPN result is already presented in Fig. \ref{fig:ana}. We show the results of AD and GBT formalisms in both metrics, while the PPN result only appears in the isotropic coordinate, since the effective Newtonian potential is better defined in the isotropic coordinate.

Let's first come to the choice of coordinate system. From Fig. \ref{fig:ad}, the AD formalism yields almost identical results in the two coordinate system, except for the upper triangle region in the negative-mass branch, which corresponds to a relatively large $|q/b_-| > 10^{-2}$ and a relatively small $|m/b_-| < 10^{-2.5}$. We comment that the upper triangle region is exactly near the ``critical line'', where the absolute value of $\alpha$ approaches $0$ and the machine error may be important, as that in the analytical case, Fig. \ref{fig:ana}. Therefore, the relative error in this region is naturally inaccurancy. Thus we conclude that the AD formalism is insensitive to the coordinate choice in most of the parameter space.

On the other hand, one may directly see from Fig. \ref{fig:gbt} that the GBT method is much more accurate in the isotropic coordinate. There seems to be a specific region (the deep blue line in the upper channel of Fig. \ref{fig:gbt}, corresponding to $q/b_- \simeq m / b_-$), where the relative error is highly small in the original coordinate. Unfortunately, the $q = m$ case would result in a apparent divergent scalar field $\phi$ \eqref{bronnikov} and a divergent mass \eqref{eq:Mpm}, so the parameter space may not be physically relevant. Thus we conclude that it would be better to apply GBT formalism in the isotropic coordinate in our exemplified EBWH case. 

\begin{figure}[ht]
    \centering
    \includegraphics[width=0.96\linewidth]{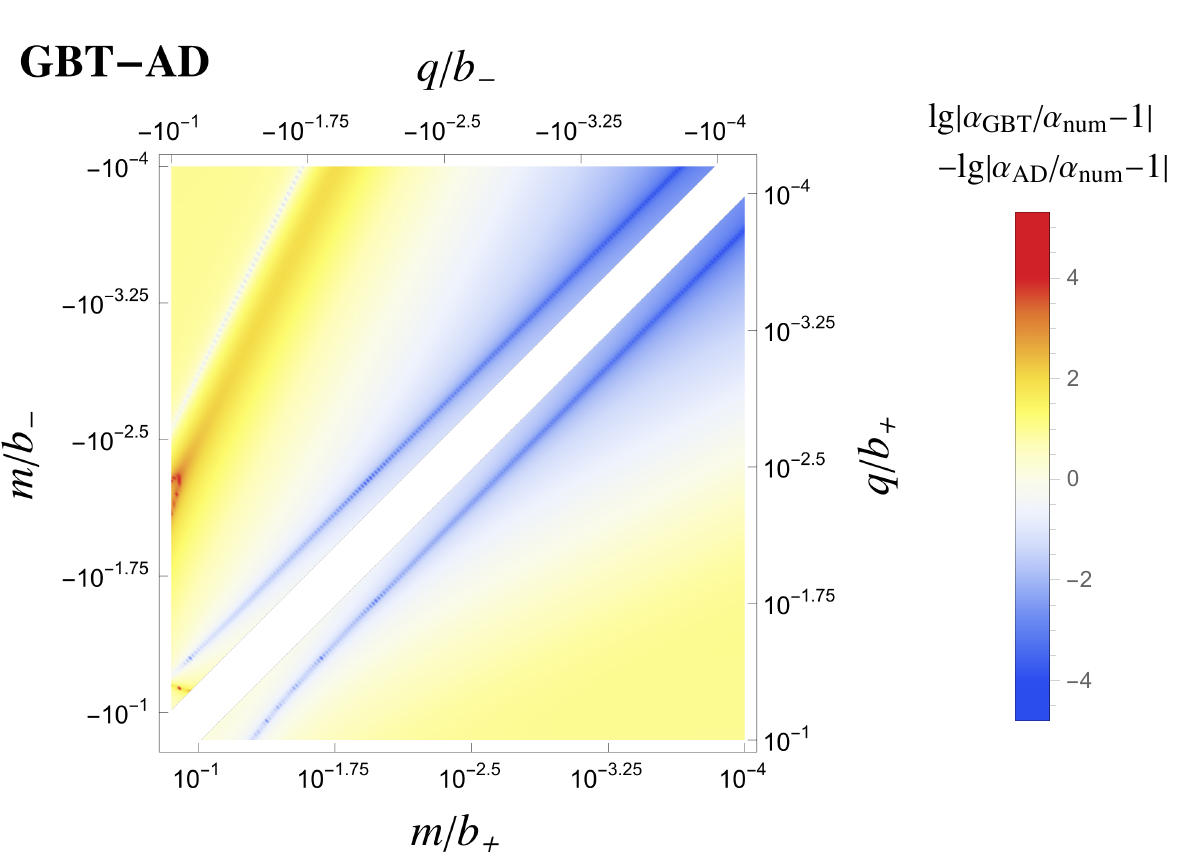}
    \includegraphics[width=0.96\linewidth]{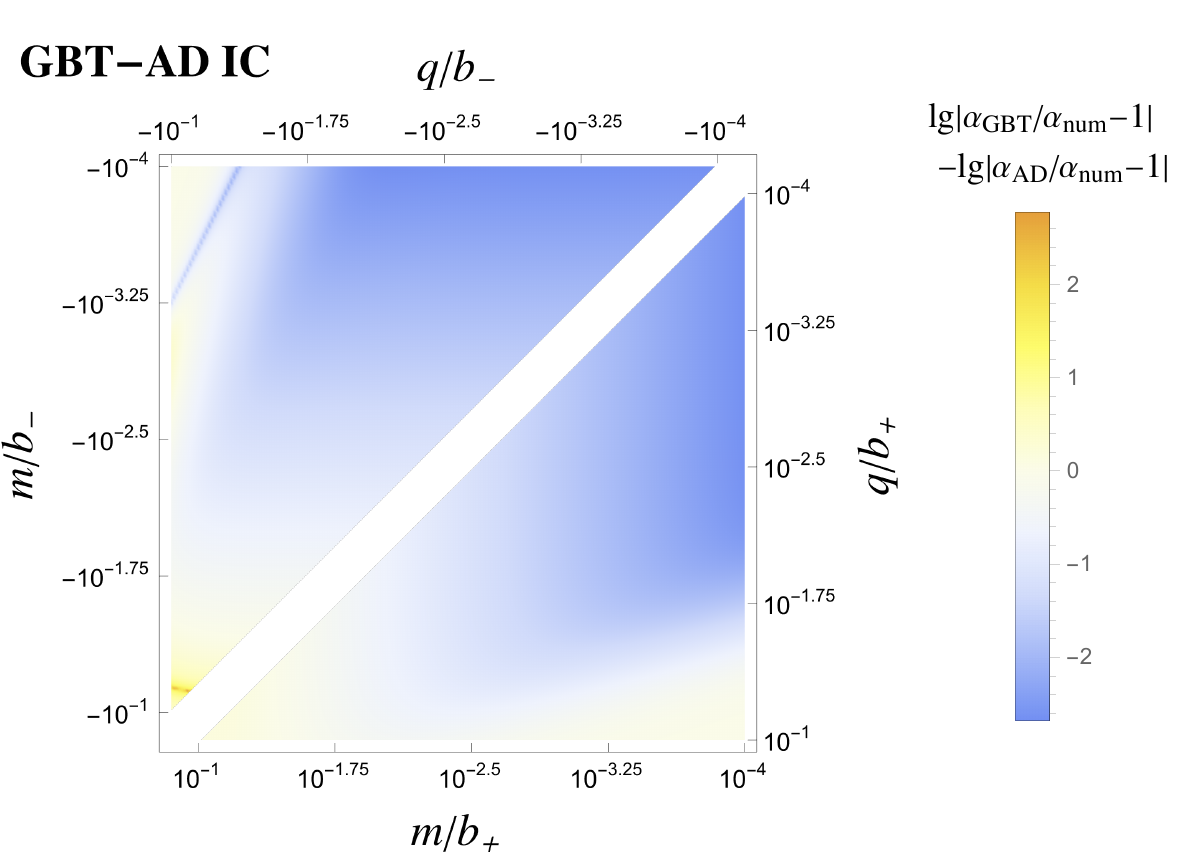}
    \caption{Comparison between the relative error of GBT and that of AD formalism for both sides of EBWH. Cool/warm color regions are where GBT/AD results are more accurate.}
    \label{fig:adgbt}
\end{figure}

Now let us come to the relative error in different methods. From Fig. \ref{fig:ana}, \ref{fig:ad} and \ref{fig:gbt}, we see that for the case of EBWH, the PPN method is the most precise one. For the reader's convenience, we depict the difference of relative errors between AD and GBT formalisms in Fig. \ref{fig:adgbt}. It is clear that, in original coordinates, the AD formalism is more precise for large $|q/b|$ and $|m/b|$, while the GBT formalism is more accurate in weaker field limit. However, in isotropic coordinates, the GBT formalism is generally more accurate than AD formalism, except for a minor parameter space.

A nice property about GBT formalism is that the weak-field condition is better satisfied. We may naively understand the result by comparing the AD and GBT results to the analytic results. The leading deviation between the AD result \eqref{eq:alphaADr}, \eqref{eq:alphaADl} and the analytic result \eqref{eq:alphaana} occurs at the $m^2/b^2_{\pm}$ term, whose coefficients differ by a proportion
\begin{equation}
    1 - \left( \frac{24}{\pi} - 8 + \frac{15}{4}\pi \right)/ \left( \frac{15\pi}{4} \right) \simeq 3\% ~.
\end{equation}
The leading deviation between the GBT result \eqref{eq:alphagbtr}, \eqref{eq:alphagbtl} and the analytic result \eqref{eq:alphaana} occurs at the $m^3/b_{\pm}^3$ term. In other words, the GBT formalism has no error at the $m^2/b^2_\pm$ order, while the AD formalism starts to show error.

We conclude that, for the EBWH case, the PPN formalism is the most accurate one, while the accuracy of AD and GBT formalisms depend on the parameter space. However, we can always turn to the isotropic coordinate, and then the GBT formalism becomes more precise than the AD formalism in most of the parameter space.

\section{Conclusion and Outlook}
\label{sec:conclusion}
The higher-order effect in gravitational lensing may be important to distinguish compact objects such as black holes and wormholes. For complicated metrics, the deflection angle may be hard to evaluate, and the approximation methods are thus implemented. Although different approximation methods give the same results on the leading order, their predictions on higher order are different, so it is important to clarify which methods should be more precise. In this work, we study the gravitational lensing effect in the EBWH wormhole under the weak-field limit to higher order. We evaluate the deflection angle to the $1/b^4_\pm$ order, using the AD formalism, the GBT formalism, and the PPN formalism, respectively. By a direct comparison to the numerical result, we find that the PPN formalism provides the most accurate result. Generically, the GBT formalism is more accurate in the weaker field limit, while the AD formalism becomes more precise in the presence of a stronger field. We also find that AD formalism is insensitive to the choice of coordinate, while the accuracy of the GBT formalism depends highly on the coordinate system. Combining all the above result, we find that the improved GBT formalism \cite{Ono:2017pie} in the isotropic coordinate gives the most accurate approximation, compared to GBT in the ordinary coordinate and the AD formalism.

Although in our case the PPN formalism yields the best result, it might be complicated to figure out all PPN parameters for sophisticated wormhole solutions. Therefore, it is still valuable to estimate the deflection angle using other formalisms at a higher order. In the EBWH case, we find that the improved GBT formalism in the isotropic coordinate is a good approximation, whose error appears at the $1/b^3$ order, thus we may attempt to apply this formula for complicated wormhole metrics, if the PPN formalism is too difficult to apply. However, since we only considered the EBWH wormhole, it would be too early to judge which formalism is better from our result. It would be interesting to compare results from different formalisms in a more generic wormhole metric and decide which formalism should be better.

It is also interesting to improve the formalism, based on the current result. For example, the AD formalism gives different predictions in the original coordinate \eqref{eq:ebsol} and isotropic coordinate \eqref{eq:isodef}. The difference may arise from the PMS, where the condition $d\alpha /d\lambda$ is different for different metrics. Would it be possible to improve the accuracy of AD formalism by changing PMS to some other principles that are coordinate-free? Besides, in the GBT formalism, the lens is simply treated as a point mass, which has no influence on the spacetime topology. However, the wormhole is a geometric structure whose existence may greatly change the spacetime topology. We may study if the topological structure of the wormhole shall be taken into account in the GBT formalism, and if the change could improve the precision of the method.

Finally, to distinguish wormholes and other compact objects, we need to evaluate astrophysical observables such as magnification and event rate. It is possible that, although the deflection angle differs for different compact objects, the resulting observable signals are still highly degenerate, whose difference is so small and below the resolution of current experiments. Thus, it is important to extend our result to astrophysical observables in concrete models, in the future.

\begin{acknowledgments}
We thank Pedro Cunha, Chunshan Lin, Lei-hua Liu, Wen-tao Luo, Xin Ren, Naoki Tsukamoto, and Yuhang Zhu for stimulating discussions. We especially thank Prof. Yi Wang for his suggestions and careful proofreading. We also thank the anonymous referee, whose valuable suggestions greatly improves the quality of our work. H.H. is grateful for support by the National Natural Science Foundation of China (NSFC)
Grant No. 12205123 and by the Sino-German (CSC-DAAD) Postdoc Scholarship Program,2021 (57575640). M.Z. is supported by Grant No. UMO 2018/30/Q/ST9/00795 from the National Science Centre, Poland.
\end{acknowledgments}

\appendix

\section{Isotropic coordinate for the EBWH}
\label{app:isotropic}

We derive the form of the isotropic coordinate for the EBWH. We start with the massless case, whose metric is
\begin{equation}
    ds^2 = - dt^2 + dr^2 + (r^2 + q^2) d\Omega_2^2 ~.
\end{equation}
Compared to \eqref{eq:isodef}, we see that $A(l) = 1$, and we have the following:
\begin{equation}
    \sqrt{B(l)} dl = dr ~,~ B(l) l^2 = r^2 + q^2 ~.
\end{equation}
For convenience we take a specific branch of solution
\begin{equation}
   2l = \sqrt{r^2 + q^2} + r ~\to~ r = l- \frac{ q^2}{4l} ~.
\end{equation}
The corresponding metric is
\begin{equation}
   ds^2 = -dt^2 + \Big(1+\frac{q^2}{4l^2}\Big)^2 \left( dl^2 + l^2 d\Omega_2^2 \right) ~.
\end{equation}

For the generic EBWH, we have
\begin{equation}
    A(l) = h(r) ~,~ B(l) dl^2 = \frac{dr^2}{h(r)} ~,~ B(l) l^2 = R^2 (r) ~.
\end{equation}
We begin with the following relation:
\begin{equation}
    \frac{dl^2}{l^2} = \frac{dr^2}{h(r)R^2(r)} ~\to~ \frac{dl}{l} = \pm \frac{dr}{\sqrt{r^2 + q^2 - m^2}} ~.
\end{equation}
Similarly we take a special branch of solution
\begin{equation}
   2l = \sqrt{r^2 + q^2 - m^2} + r ~\to~ r = l- \frac{ q^2-m^2}{4l} ~.
\end{equation}
We will use the identity
\begin{equation}
    2\arctan(x^2+\sqrt{x^2+1})=\frac{\pi}{2}+\arctan x,
\end{equation}
and define a dimensionless parameter 
\begin{equation}
    \gamma \equiv \frac{m}{\sqrt{q^2 - m^2}} ~,
\end{equation}
such that
\begin{equation}
    A = \exp \left[ \gamma \left(\pi - 4 \arctan\frac{2l\gamma}{m} \right) \right] ~,
\end{equation}
\begin{equation}
     B = \frac{1}{A}  \Big(1+\frac{q^2-m^2}{4l^2}\Big)^2  ~.
\end{equation}

\bibliography{apslens}% Produces the bibliography via BibTeX.
\bibliographystyle{utphys}

\end{document}